\documentclass[aps,prl,reprint,longbibliography,superscriptaddress]{revtex4-2}
\usepackage{mathrsfs}
\usepackage{amsmath,gensymb}
\usepackage{amsfonts}
\usepackage{amssymb}
\usepackage{amsthm}
\usepackage{graphicx}
\usepackage{natbib}
\usepackage{color}
\usepackage{hyperref}
\usepackage{bm}
\usepackage[caption=false]{subfig}
\usepackage{verbatim}
\usepackage[normalem]{ulem}
\usepackage{siunitx}

\begin{document}



\title{Multiterminal Quasiballistic Josephson Effect in Monocrystalline Gold}

\author{K.B. Polevoy}
\affiliation{Moscow Institute of Physics and Technology, Dolgoprudny, 141700 Moscow region, Russia}
\affiliation{All-Russian Research Institute of Automatics n.a. N.L. Dukhov (VNIIA),
127030, Moscow, Russia}

\author{G.A. Bobkov}
\affiliation{Moscow Institute of Physics and Technology, Dolgoprudny, 141700 Moscow region, Russia}

\author{D.S. Kalashnikov}
\affiliation{Moscow Institute of Physics and Technology, Dolgoprudny, 141700 Moscow region, Russia}



\author{A.G. Shishkin}
\affiliation{Moscow Institute of Physics and Technology, Dolgoprudny, 141700 Moscow region, Russia}

\author{I.V. Trofimov}
\affiliation{Institute of Nanotechnology of Microelectronics of RAS, 119991, Moscow, Russia}

\author{A.M. Bobkov}
\affiliation{Moscow Institute of Physics and Technology, Dolgoprudny, 141700 Moscow region, Russia}

\author{E.A. Sedov}
\affiliation{Moscow Institute of Physics and Technology, Dolgoprudny, 141700 Moscow region, Russia}

\author{E.A. Krivko}
\affiliation{Institute of Nanotechnology of Microelectronics of RAS, 119991, Moscow, Russia}
\affiliation{Moscow Institute of Physics and Technology, Dolgoprudny, 141700 Moscow region, Russia}

\author{M.A. Tarkhov}
\affiliation{Institute of Nanotechnology of Microelectronics of RAS, 119991, Moscow, Russia}
\affiliation{Moscow Institute of Physics and Technology, Dolgoprudny, 141700 Moscow region, Russia}

\author{I.V. Bobkova}
\affiliation{Moscow Institute of Physics and Technology, Dolgoprudny, 141700 Moscow region, Russia}
\affiliation{HSE University, 101000 Moscow, Russia}

\author{V.~S.~Stolyarov}
\affiliation{Moscow Institute of Physics and Technology, Dolgoprudny, 141700 Moscow region, Russia}
\affiliation{All-Russian Research Institute of Automatics n.a. N.L. Dukhov (VNIIA),
127030, Moscow, Russia}


\begin{abstract}

We report on the realization of a planar, quasiballistic Josephson junction array using a Au micron-sized single crystal. The system exhibits a nonlocal, multiterminal Josephson effect, where the supercurrent between any two superconducting leads is governed by the phase coherence across the entire crystal. Key evidence includes a nonmonotonic dependence of the critical current on junction length and magnetic interference patterns with periods corresponding to the shared normal-metal area. Nonlocal transport measurements further confirm that the supercurrent between two electrodes depends on the phase configuration of all the others. Our results, supported by a developed theoretical model, establish a platform for exploring complex superconducting phenomena in multiterminal ballistic systems.

\end{abstract}

\maketitle

{\it Introduction.}---The Josephson effect continues to inspire fundamental research and technological applications since its discovery in 1962 \cite{Josephson1962}. Planar Josephson junction technology offers a promising route to miniaturize weak-link regions to 100--200 nm while maintaining high critical currents $I_c$ and characteristic voltages $V_c = I_c R_N$. Typically, such Josephson junctions comprise a normal metal (N) nanowire or film with two superconductor electrodes (S) deposited on top \cite{hoss2000multiple,jung2011superconducting,wang2009proximity,sotnichuk2022long,Beasley1981,nakano1987magnetic,nakano1986nb,nakano1987fabrication,ichikawa1994current,dubos2001josephson,savin2004cold,garcia2009josephson,golikova2013critical,golikova2014nonlocal,crosser2008nonequilibrium,Cuevas2008,morpurgo1998hot,baryshev1989theoretical,skryabina2024anomalous}, forming complex SN/N/SN Josephson junctions (JJs) in bridge geometry. Most experimental and theoretical studies \cite{Golubov2004,Soloviev2021,Ruzhickiy2023,Bosboom2021,Marychev2020,2024_Bakurskiy,Polevoy_2026,Vodolazov_2026,Bakurskiy_2026} address the diffusive regime, where the mean free path $l$ is less than the superconducting coherence length $\xi_N$.

Planar ballistic JJs, enabling Josephson coupling over micron-scale distances, provide a unique platform for studying proximity-induced superconductivity in the ballistic regime. Up to now, the ballistic JJs have been mainly implemented and studied  experimentally in heterostructures of
graphene \cite{Calado2015,BenShalom2016,Allen2016,Kumaravadivel2016,Borzenets2016,Zhu2018}. Moreover, in more complex S/N/S structures, where a normal metal is proximitized by several superconducting electrodes, novel effects come into play. In particular, the energy spectrum of a multi-terminal Josephson device can emulate a band structure supporting Weyl points, where the superconducting phases play the role of quasi-momenta \cite{vanHeck2014,Riwar2016,Gavensky2023}. Thus, multi-terminal Josephson junctions are expected to be a platform for designing and studying the topological matter of arbitrary dimensions. Several groups have already implemented mesoscopic JJs with three or four terminals in diffusive metallic junctions \cite{Pfeffer2014,Strambini2016,Vischi2017}, hybrid semiconductor-superconductor heterostructures \cite{Cohen2018,Graziano2020,Pankratova2020,Graziano2022}, and topological materials \cite{Kozler2023}. Ballistic SN/N/SN JJs, where the superconducting electrodes are deposited on top of the normal crystal, are particularly attractive for multiterminal systems, as they facilitate efficient coupling between leads separated by several microns, as demonstrated in graphene \cite{Draelos2019,Arnault2021,Huang2022}. 

In this Letter, we present a combined experimental and theoretical study of planar, quasi-ballistic multi-terminal JJs comprising Al superconducting leads on a gold single crystal. We investigate supercurrent flow as a function of temperature and magnetic field and develop a microscopic quasiclassical theory for quasi-ballistic transport in multi-terminal planar S/N/S structures. The theory agrees quantitatively with experiments, confirming the realization of a planar multi-terminal Josephson system in the quasi-ballistic regime. 

\begin{figure*}[t]    
 \includegraphics[width=170mm]{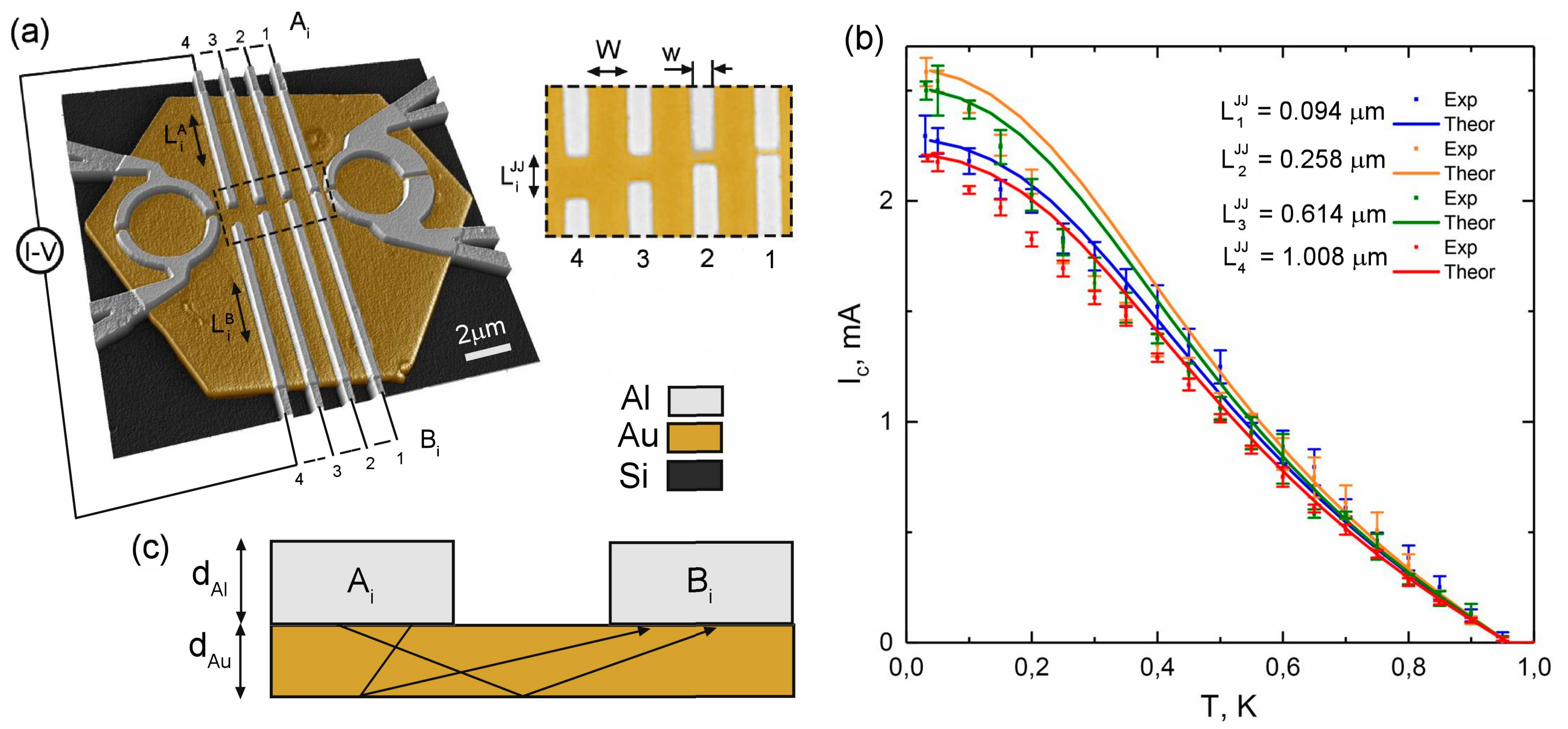}
\caption{(a) False-colored SEM image of the Josephson junction array with Al electrodes (labeled $A_1$--$A_4$, $B_1$--$B_4$) on a Au crystal. Inset: measurement configuration schematic. (b) Critical current $I_c^{A_i-B_i}$ versus temperature: experimental data (dots) and theoretical fits (lines) with interface transmission $D = 0.03$. (c)  Model for Josephson current calculation in a single junction.}
  \label{fig:setup}
\end{figure*}

{\it Sample fabrication and characterization.}--- We fabricate ballistic Josephson junctions by depositing aluminum ($d_{\text{Al}} = \SI{175}{\nano\meter}$) through an electron-beam lithography mask onto a mechanically transferred monocrystalline Au flake ($d_{\text{Au}} = \SI{80}{\nano\meter}$, lateral size \SI{14}{\micro\meter}) on a $\text{SiO}_2$/Si substrate. The monocrystalline Au flakes were synthesized using a modified Brust-Schiffrin method followed by thermolysis, as described in Ref.~\cite{brust1994synthesis,radha2011real,menabde2022near}. The Au flake was transferred onto the substrate using an all-dry viscoelastic stamping technique with a polyvinyl chloride/polydimethylsiloxane (PVC/PDMS) stamp~\cite{onodera2022all}. Figure~\ref{fig:setup}(a) shows a false-colored SEM image of the device, with eight Al electrodes (labeled $A_1$--$A_4$ and $B_1$--$B_4$) contacting the Au crystal for standard four-probe dc (direct current) measurements. The inset schematically shows the additional four-probe dc measurement configuration. Transport measurements are performed in a dilution refrigerator using filtered dc lines. From resistivity measurements, $\rho(\SI{300}{\kelvin}) = \SI[inter-unit-product = \cdot]{20.70}{\nano\ohm\meter}$ and $\rho(\SI{4}{\kelvin}) = \SI[inter-unit-product = \cdot]{1.35}{\nano\ohm\meter}$, we extract a low-temperature electron mean free path, $l = \SI{680}{\nano\meter}$. The junction lengths are $L^{JJ}_{1} = \SI{94}{\nano\meter}$, $L^{JJ}_{2} = \SI{258}{\nano\meter}$, $L^{JJ}_{3} = \SI{614}{\nano\meter}$, and $L^{JJ}_{4} = \SI{1008}{\nano\meter}$. Each lead has a length of $\sim\SI{7}{\micro\meter}$ and a width $w = \SI{570}{\nano\meter}$, with an inter-lead spacing of $W = \SI{1}{\micro\meter}$.

{\it Zero-field critical currents.}---
Fig.~\ref{fig:setup}(b) displays the temperature dependence of the critical current $I_c$ for the $A_i$--$B_i$ junctions ($i=1$--$4$). Strictly speaking, the experimentally measured quantity is the switching current, which is systematically smaller than the true critical current. In our case, however, the difference is negligible owing to the smallness of thermal fluctuations and capacitive effects. We therefore use the term critical current throughout. Strikingly, $I_c$ depends nonmonotonically on interlayer length $L_{i}^{JJ}$: the critical currents of the two middle JJs $I_c^{A_2-B_2}(T=200 \mathrm {mK}) \approx I_c^{A_3-B_3}(T=200 \mathrm {mK}) = I_c^m$  are very close to each other and exceed the currents of the two outer JJs $I_c^{A_1-B_1}(T=200 \mathrm {mK}) \approx I_c^{A_4-B_4}(T=200 \mathrm {mK}) = I_c^o$, which are also close to each other and are smaller than $I_c^m$ by a factor of $I_c^o/I_c^m \approx 7/8$. This occurs despite the monotonic increase of $L_{i}^{JJ}$ from $A_1$--$B_1$ to $A_4$--$B_4$, indicating that the four JJs are not independent. Using the Au Fermi velocity $v_F=\SI{1.4e6}{\meter\per\second}$ and Al critical temperature $T_c = \SI{0.95}{\kelvin}$, the ballistic coherence length is $\xi = v_F/2\pi T_c = \SI{1.7}{\micro\meter}$, exceeding all $L_{i}^{JJ}$ and comparable to inter-electrode distances. Thus, we come to the conclusion that Josephson coupling occurs between all pairs of superconducting leads.
 This hypothesis is fully confirmed by our theoretical calculations, which reproduce the ratios between the different $I_c^{A_i-B_i}(T=0,B=0)$ without any fitting parameters, thereby explaining the absence of a monotonic dependence of the critical current amplitude on the length of the weak-link region.

Existing theories for planar SN/N/SN structures assume either diffusive transport or purely ballistic trajectories, neither of which is applicable here. The Au crystal thickness $d$ is less than the mean free path, and its width significantly exceeds the electrode dimensions. Moreover, no direct ballistic trajectories connect the leads; only trajectories involving reflections contribute [Fig.~\ref{fig:setup}(c)]. Previous ballistic junction models \cite{Kuprianov1986,Kuprianov1988} assume translationally invariant infinite leads, which are inadequate for multiterminal structures. We develop a theoretical approach for strongly coupled planar Josephson structures on large crystals, detailed in a companion work \cite{Bobkov2025_joint}. Here, we apply it to our Al/Au/Al junctions.

If the thickness of the normal metal is large compared to the mean free path and the coherence length, that is $d \gtrsim (\xi,l)$, then the main contribution to the Josephson current is given by the impurity scattering in the normal layer. If the opposite case $d \ll (\xi,l)$ occurs, the main contribution to the current is provided via the mechanism of reflections from the surfaces of the normal crystal. Our experiment corresponds to the second case, and for this reason, we focus on this limit, while a more complete theory is discussed in the companion paper \cite{Bobkov2025_joint}. The central idea behind calculating the Josephson current in such a complex multiterminal system, where every superconducting electrode is Josephson-coupled to all others, is to partition each electrode into small elements and first determine the Josephson current between any two elements $S_1$ and $S_2$ belonging to two arbitrary electrodes.
For this inter-element current, the following approximate expression was derived in Ref.~\cite{Bobkov2025_joint}: 
\begin{align}
i_{S_1\to S_2}=&T \sum \limits_\omega\frac{D^2e|\Delta|^2 dS_{S1} dS_{S2} N_F \lambda^2 v_F\sin(\phi_2-\phi_1)}{8\pi (|\Delta|^2+\omega^2)} \times  \nonumber \\
&\frac{2}{\lambda^2(L^2+4d^2)}(2e^{-\lambda\sqrt{L^2+4d^2}}+e^{-\lambda (L+2d)})
\label{eq:current_diffusive_approx}
\end{align}
where $dS_{S_1,S_2}$ are the areas of the two superconducting elements, $N_F$ is the density of states at the Fermi level in the normal layer, $|\Delta|$ is the absolute value of the superconducting order parameter (assumed identical in all leads), $\phi_{1,2}$ are the corresponding order-parameter phases, $L$ is the distance between the elements, and $\lambda = 2\omega/v_F$ with the Matsubara frequency $\omega = 2\pi T (2n+1)$ and integer $n$.
In this model, the multiple diffusive reflections from the N layer surfaces are disregarded. This approximation works well if $\xi<d<l$ or $L \lesssim 2d$. The first of the conditions is not applicable to the current experiment, but the second condition is fulfilled for part of the JJs. 

To calculate the net current $I^{A_i-B_i}$ corresponding to the current flow from $A_i$ to $B_i$ electrodes, we first calculate the critical currents $i^{A(B)_i-A(B)_j}$ between all possible pairs of $A(B)_i-A(B)_j$ and $A_i-B_j$ by summing the currents over all pairs of small superconducting elements belonging to the corresponding leads:
\begin{align}
    i^{A(B)_i-A(B)_j}=\sum \limits_{S_1 \in A(B)_i,S_2 \in A(B)_j} i_{S_1\to S_2}.
    \label{eq:I_total}
\end{align}
Then the superconducting phases are set at two leads through which the current is passed, and the other phases are determined from the condition of current conservation \cite{Bobkov2025_joint}.

In Fig.~\ref{fig:setup}(b), the calculated $I_c^{A_i-B_i}$ JJs for $i=1-4$ are superimposed on the experimental data. The model provides very good quantitative agreement for the low-temperature values of all four critical currents $I_c^{A_i-B_i}$. The only fitting parameter is the transparency of the S/N interface $D$. However, it only controls the absolute value of the critical current amplitude, while the ratio between different $I_c^{A_i-B_i}$ with $I_c^o/I_c^m \approx 7/8$ is reproduced without any fitting parameters. 

Whereas the quantitative value of the ratio between different $I_c^{A_i-B_i}$ depends on the specific geometry of the system and can only be obtained by direct calculation, the qualitative origin of this nontrivial behavior---different critical currents whose amplitudes are nevertheless unrelated to the length of the individual weak link---can be understood as follows.
In the quasi-ballistic system under study, all partial currents $i_c^{A_i-B_i}$ flowing through individual electrode pairs $A_i$--$B_i$ are close in amplitude~\cite{Bobkov2025_joint}, simply because the length of each weak link does not exceed the coherence length, which corresponds to the short-junction limit.
At the same time, the total current $I_c^{A_i-B_i}$ through a given electrode pair receives contributions not only from the ``direct path'' $i_c^{A_i-B_i}$, but also from all ``indirect paths'' via intermediate electrodes.
For a quasi-ballistic system, this latter contribution is by no means small.
The outer electrode pairs $A_{1,4}$--$B_{1,4}$ have fewer such indirect paths than the middle pairs $A_{2,3}$--$B_{2,3}$, which explains the systematic character of the relation $I_c^{\mathrm{o}} < I_c^{\mathrm{m}}$.
This relation will hold for arbitrary weak-link lengths $L_i^{\mathrm{JJ}}$, provided they remain shorter than the coherence length.

To definitively rule out alternative interpretations, such as random sample-to-sample variations of transparencies for a series of junctions, and to demonstrate that the observed behavior is intrinsic to the quasi-ballistic system, we have performed additional control experiments. Devices with identical contact geometries were fabricated using conventional diffusive gold. The results were dramatically different: the critical current of the shortest junction at $T\to 0$ was $\sim 1.9 \mu$A; for the next-longest junction, the critical current dropped to $\lesssim 0.2 \mu$A; for the third and fourth junctions, no superconducting current was observed at all. 
This unambiguously indicates a conventional exponential decay of the critical current with junction length, characteristic of the diffusive regime, and stands in sharp contrast to the data obtained on our quasi-ballistic structure. 

The overall nearly linear temperature dependence of the critical current is also reproduced, although the quantitative agreement between theory and experiment at intermediate temperatures is not ideal. In the experiment, the deviation of the $I_c(T)$ from linearity occurs at lower temperatures than the theory suggests. In our opinion, this is due to the fact that, in the experiment, the interlayer lengths of most JJs are of the same order as the mean free path $l$; therefore, the ballistic limit condition is not fully satisfied. That is, in addition to diffusive reflection from the surfaces of the N layer, there is also a certain amount of reflection from impurities in the N layer itself, which changes the temperature dependence of the coherence length. 

\begin{figure}[t]
\includegraphics[width=84mm]{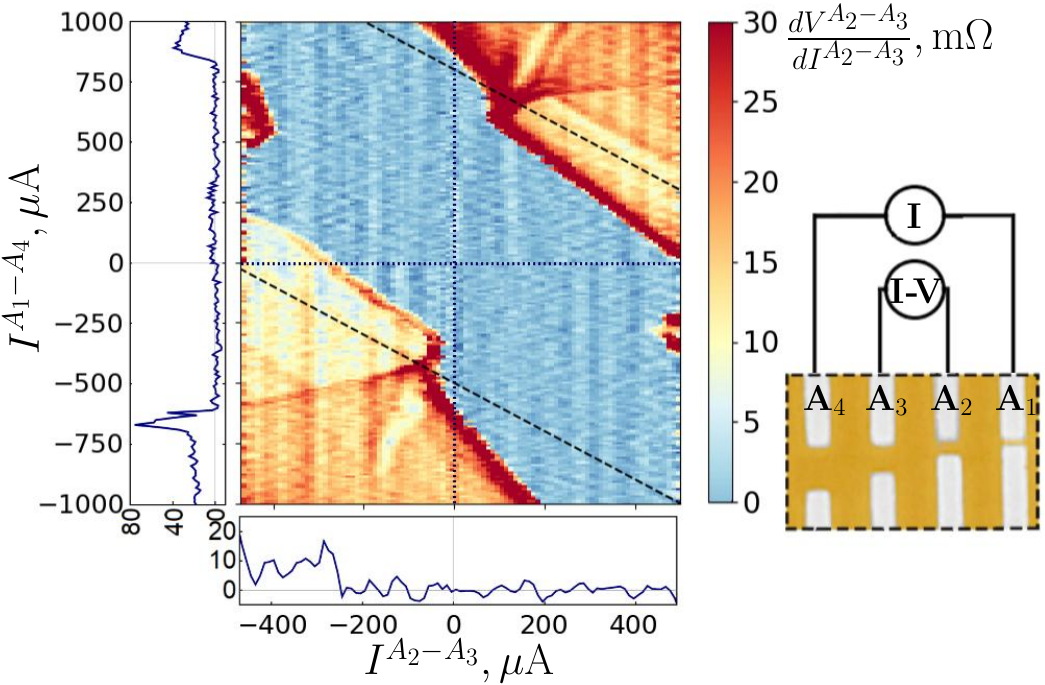}
\caption{Differential resistance $dV/dI$ measured at the $A_2$-$A_3$ junction as a function of the injected currents $I^{A_2-A_3}$ and $I^{A_1-A_4}$ at $T = 0.7$ K.
The blue region marks the superconducting state where $dV/dI = 0$. The black dotted lines indicate the boundaries expected for a simple series chain of three independent junctions, with slopes of $-1$.
The line profiles $dV/dI (I^{A_2-A_3})|_{I^{A_1-A_4}=0}$ (horizontal) and $dV/dI (I^{A_1-A_4})|_{I^{A_2-A_3}=0}$ (vertical) are also shown.}
 \label{fig:multiterminal}
\end{figure}

{\it Multiterminal transport.}---Fig.~\ref{fig:multiterminal} shows the resistance $dV/dI$ measured at the $A_2$--$A_3$ junction when the current $I^{A_2-A_3}$ is applied from $A_2$ to $A_3$ and the current $I^{A_1-A_4}$ is applied from $A_1$ to $A_4$ simultaneously. Line profiles $dV/dI (I^{A_2-A_3})|_{I^{A_1-A_4}=0}$ and $dV/dI (I^{A_1-A_4})|_{I^{A_2-A_3}=0}$ are also shown. These data provide evidence of the multiterminal behavior of our system, demonstrating that $I^{A_2-A_3}$ depends not only on the phase difference at the $A_2$--$A_3$ junction but also on the phases at other superconducting electrodes.

Indeed, if $I^{A_i-A_j}$ were determined solely by the phase difference at the $A_i$--$A_j$ junction between nearest-neighbor electrodes, the system of $A_1$, $A_2$, $A_3$, and $A_4$ electrodes on top of the Au crystal could be viewed as a chain of three junctions in series: $A_1$--$A_2$, $A_2$--$A_3$, and $A_3$--$A_4$. In that case, the boundary of the blue region would be given by $I^{A_2-A_3}+I^{A_1-A_4}= \pm i_c^{A_2-A_3}$, implying slopes of $-1$ (black dotted lines in Fig.~\ref{fig:multiterminal}). The observed slopes, however, are $\approx -1.7$, indicating that, for example, $I_c^{A_1-A_3} \neq 0$ and, consequently, $I^{A_2-A_3}$ depends on the phases at all electrodes.

Furthermore, if the system consisted merely of three junctions in series ($A_1$--$A_2$, $A_2$--$A_3$, and $A_3$--$A_4$), the cut along $I_{14}=0$ would simply give the $I$--$V$ characteristic of the isolated $A_2$--$A_3$ junction; the critical current extracted from this cut would thus equal $i_c^{A_2-A_3}$. The cut along $I_{23}=0$ would yield the differential resistance of the $A_2$--$A_3$ junction with current passing through the entire chain. In this scenario, phase slips, capacitive effects, or heating could make the extracted critical current smaller than (or, ideally, equal to) $i_c^{A_2-A_3}$. Instead, the experiment yields a value approximately a factor of two larger (see horizontal and vertical line profiles in Fig.~\ref{fig:multiterminal}). Both findings confirm the multiterminal nature of the system.

\begin{figure}[t]
\includegraphics[width=84mm]{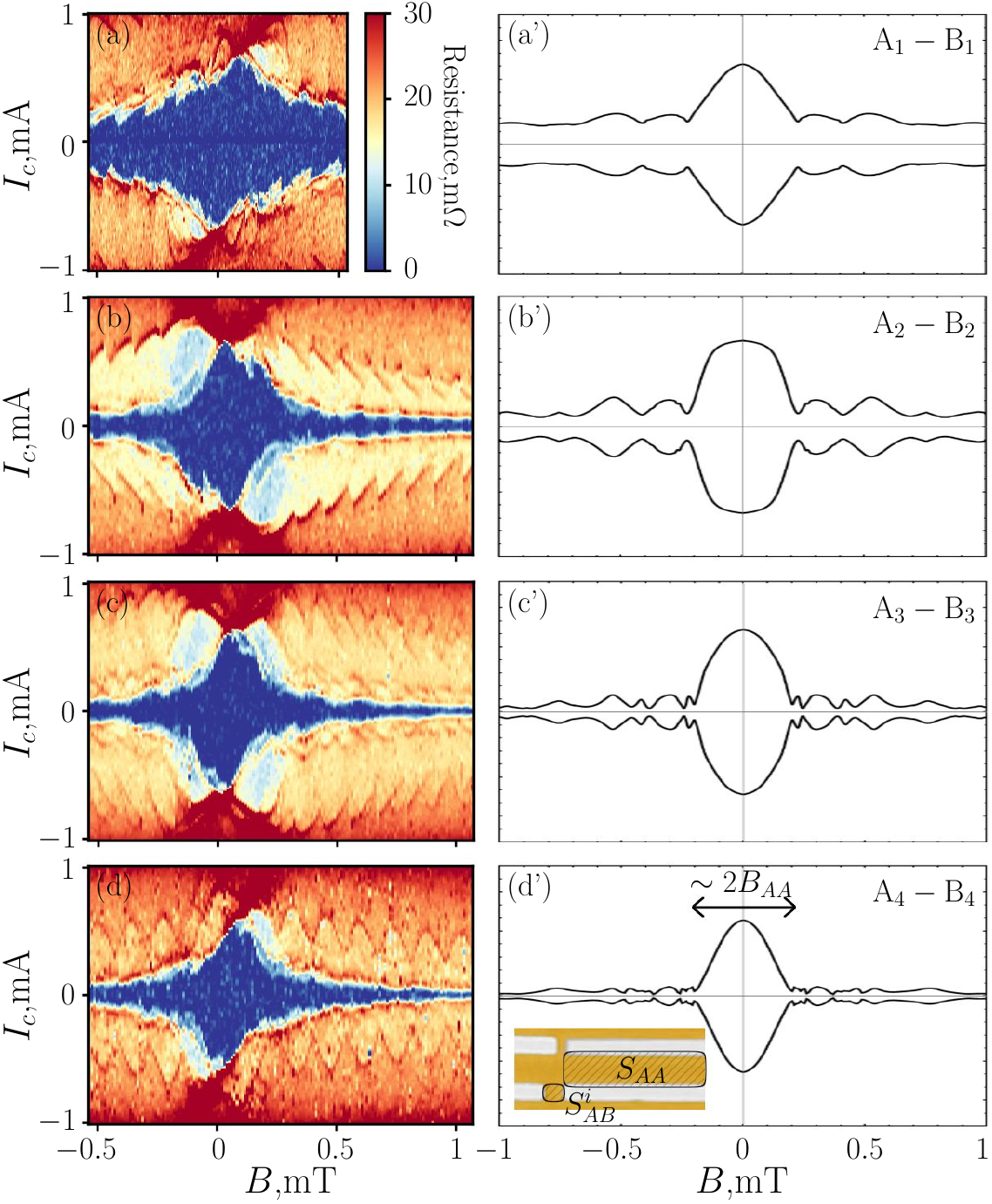}
\caption{(a)-(d) Experimental data for $I_c^{A_i-B_i}$ (dark blue region boundaries) versus perpendicular magnetic field. (a')-(d') Theoretical results for $I_c^{A_i-B_i}$ versus perpendicular magnetic field. $T=\SI{0.7}{\kelvin}$. The inset in (d') schematically defines the $A_i-B_i$ junction area $S_{AB}$ and $A_i-A_j$ junction area $S_{AA}$.}
 \label{fig:current_field_exp}
\end{figure}

{\it Magnetic field response.}---The experimental data for the dependence of $I_c^{A_i-B_i}$ on the applied magnetic field $B$ are presented in Figs.~\ref{fig:current_field_exp}(a)-(d).  First of all, one can see that the period of the Fraunhofer-like oscillations of $I_c^{A_i-B_i}$ is approximately two orders of magnitude smaller than expected for an independent $A_i-B_i$ JJ $B_{AB} = \Phi_0 / S_{AB}$, where $S_{AB}=\rm w L_{i}^{JJ} $ is the area of the $A_i-B_i$ weak link (see inset to Fig.~\ref{fig:current_field_exp}(d')), and $\Phi_0 = \pi \hbar /e$ is the flux quantum. $B_{AB}=3\sim30$mT for $i\in [1,...4]$. The observed period approximately corresponds to the large area $S_{AA(BB)}=L_i^{A(B)}W$ of the normal interlayers of the $A(B)_i-A(B)_j$ JJs between the nearest neighbors $i$ and $j$. It once again proves our main statement that, due to the quasi-ballistic regime of transport corresponding to the large coherence length $\xi \approx 1.7\mu$m, the Josephson coupling occurs between all the superconducting elements in the system. The asymmetry between the positive and negative maxima of $I_c$ can be explained by the magnetic self-fields generated by the Josephson currents, which are not accounted for in our theoretical treatment, and, possibly, by magnetic vortices trapping in the sample or magnet during field sweeps.  Among other possible origins of this asymmetry, diode effects arising from high-transmission conduction channels can be significant in Josephson systems~\cite{Souto2022,Seleznev2024,Kalashnikov2025}. We believe, however, that this mechanism is unlikely to dominate in our experiment, as the estimated interfacial transparency of our junctions is too low.

The results of the calculation in the framework of the developed quasi-ballistic theory \cite{Bobkov2025_joint} are presented in Figs.~\ref{fig:current_field_exp}(a')-(d'). The theory reproduces well both the maximum current amplitude at $B=0$ and all the main features of the curves: a sharp drop in the critical current at a field of $\sim 0.2$ mT, which is the same for all JJ, followed by a weakly decaying behavior on which quasi-periodic oscillations are superimposed. The initial sharp drop at $B \sim 0.2$ mT is attributed to the first minimum of the Fraunhofer pattern of $i_c^{A(B)_i-A(B)_j}$ and is located at $B_{AA} \approx \Phi_0/(L_i^{A(B)} \rm W)$. At larger fields, the main contribution to the current is produced by $A_i-B_i$ JJs, which have a much smaller area of the normal interlayer region and a much weaker, non-oscillating dependence on the field \cite{Bobkov2025_joint}. It is important that at $B>0.2$ mT, when the contribution of the $A(B)_i-A(B)_j$ JJs is strongly suppressed, the $A_i-B_i$ JJs become practically independent. For this reason, the average critical current at $B>0.2$ mT, which corresponds to $i_c^{A_i-B_i}$, monotonically decreases with increasing $L_{i}^{JJ}$ from $i=1$ to $4$. The quasiperiodic character of the oscillations is the result of the superimposition of the Fraunhofer patterns corresponding to $A(B)_i-A(B)_j$ JJs, which correspond to slightly different $L_i^{A(B)}$. 
 

{\it Conclusions.}---We demonstrate a multi-terminal superconducting system on a Au single crystal in the quasi-ballistic regime, where the superconducting coherence length is comparable to inter-electrode distances. The nonmonotonic critical current dependence on interlayer length, distinctive magnetic field response showing a sharp regime crossover, and direct multiterminal transport measurements provide compelling evidence for strong coupling among all superconducting leads. Our results establish a platform for exploring multi-terminal Josephson effects, including artificial topological matter in metallic mesoscopic devices, where the entire Au crystal serves as the normal scattering region \cite{Bobkov2025_joint}. The developed theoretical approach for strongly coupled planar Josephson structures on large quasi-ballistic crystals agrees well with experiments, strongly supporting our interpretation.

\begin{acknowledgments}
We thank Sir A. Geim and E. Nguyen for the fruitful discussions. The work is supported by a grant RSF No 25-42-00058 \href{https://rscf.ru/project/25-42-00058/} 
{https://rscf.ru/project/25-42-00058/} (experiment) and RSF No 25-72-10127 \href{https://rscf.ru/project/25-72-10127/}
{https://rscf.ru/project/25-72-10127/}(theory)
This work was carried out using equipment from the MIPT Shared Facilities Center.
\end{acknowledgments}

\bibliography{SNSballistic}

\end{document}